\documentclass[12pt]{iopart}
\usepackage{graphicx}
\begin{document}
\title{Low energy neutrino scattering measurements at future Spallation Source facilities}
\author{R. Lazauskas$^1$, C. Volpe$^2$}

\address{$^1$ IPHC, IN2P3-CNRS/Universit\'e Louis Pasteur BP 28, F-67037 Strasbourg Cedex
2, France}
\address{$^2$ Institut de Physique Nucl\'eaire, F-91406 Orsay cedex, CNRS/IN2P3 and University of Paris-XI, France}
\ead{rimantas.lazauskas@ires.in2p3.fr,volpe@ipno.in2p3.fr}

\begin{abstract}
In the future several Spallation Source facilities will be available worldwide.
Spallation Sources produce large amount of neutrinos from decay-at-rest muons
and thus can be well adapted to accommodate state-of-the-art neutrino experiments.
In this paper low energy neutrino scattering experiments that can be performed at
such facilities are reviewed. Estimation of expected event rates are given
 for several nuclei, electrons and protons at a detector located close to the source.
A neutrino program at Spallation Sources comprises neutrino-nucleus cross
 section measurements relevant for neutrino and
 core-collapse supernova physics, electroweak tests and lepton-flavor violation searches.

\end{abstract}
\pacs{25.30.Pt,97.60.Bw,26.30.Jk,26.30.Hj}
\submitto{\JPG}
\maketitle
\normalsize
\section{Introduction}
Spallation sources can be well adapted to accommodate innovative neutrino experiments.
At such installation large amounts of neutrinos are produced from the decay of pions,
i.e. $\pi^+ \rightarrow \mu^+ + \nu_{\mu} $ and the subsequent decay of muons
$\mu^+ \rightarrow \e^+ + \nu_e + \bar{\nu}_{\mu}$; while,
due to the strong anti-pion absorption in the spallation target, very few
electron anti-neutrinos evolve via the charge-conjugate pion decay sequence
(typically at the level of 10$^{-5}$).
Spallation sources have already been used in the past to study neutrino oscillations by the LSND~\cite{Athanassopoulos:1996jb,Athanassopoulos:1997pv,Athanassopoulos:1997er} and KARMEN \cite{Armbruster:2002mp} collaborations
at LANSCE in Los Alamos and ISIS at Rutherford Appleton Laboratory respectively.
The LSND (KARMEN) experiment has used a 167 (65) tons scintillator detector located at about 17 (17.6) meters from the LANSCE (ISIS) source.
In parallel the same experimental setups have been exploited to perform a
variety of experiments including measurements of neutrino scattering on $^{12}$C~\cite{Albert:1994xs,Athanassopoulos:1997rn,Athanassopoulos:1997rm,Krakauer:1991rf,Allen:1990nr,bib:nuC12,Bodmann:1994py,Drexlin:1991gx} $^{13}$C, and $^{56}$Fe~\cite{Zeitnitz:1998qg} as well as electroweak tests (lepton flavor universality~\cite{Bodmann:1994py}, V-A structure in muon decay~\cite{Armbruster:1998qi}, Weinberg angle \cite{Auerbach:2001wg}) and lepton-flavor violation searches~\cite{Armbruster:2003pq}.
New spallation sources are being constructed, are planned or under study
including the Spallation Neutron Source (SNS) facility at Oak Ridge~\cite{sns}, the European Spallation Source (ESS) facility in Lund, the Japanese Spallation Neutron Source
(JSNS) at JPARC~\cite{future} and the SPL at CERN~\cite{Bandyopadhyay:2007kx}. They will offer the possibility to realize a low energy neutrino physics program of interest for particle physics\footnote{Note that in \cite{Conrad:2009mh} a new idea for the search of CP violation in the lepton sector is proposed, using spallation sources.}, neutrino astrophysics and nuclear physics.
\begin{figure}[t]
\begin{center}
\includegraphics[scale=0.7,angle=0]{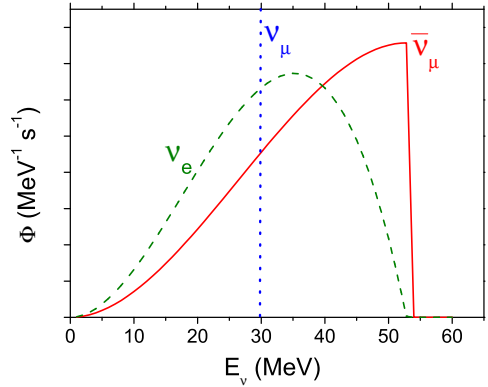}
   \caption{Neutrino fluxes from muon and pion decay-at-rest at a spallation source.}
\end{center}
   \label{fig:flux}
\end{figure}
The only technical alternative to produce controlled neutrino fluxes in the 100 MeV energy range is a low energy beta-beam facility~\cite{Volpe:2003fi}, which 
is based on the novel method of the beta-beams~\cite{Zucchelli:2002sa} exploiting the beta-decay of boosted radioactive ions.  While the main goal of
the beta-beam  is the search of CP violation in the lepton sector, the physics program of the low energy variant covers a variety of interesting
topics in nuclear physics, in the study of fundamental interactions and core-collapse supernova physics (for a review of beta-beams see~\cite{Volpe:2006in}).
An advantage of the beta-beam facility is in its capacity to produce collimated
beams of both $\nu_e$ and $\overline{\nu_e}$ species, as well as the possibility they
provide to control the average energy of the neutrino beams.
On the other hand spallation sources have significantly larger neutrino production rates.

Determination of neutrino cross sections in a broad range of energies (several hundred MeV to multi-GeV range) is currently the focus of several experiments
including Miner$\nu$a~\cite{minerva} and Sciboone~\cite{sciboone}.
Note that recently the MiniBOONE results revealed an excess of the electron-like events for the several hundred MeV neutrino energy range~\cite{AguilarArevalo:2008rc}.
The neutrino scattering experiments that can be performed at Spallation Sources, where the electron neutrino energy distribution is
described by the Michel spectrum with the endpoint at about 52.8 MeV (Figure~\ref{fig:flux}),
turns to be complementary, as they test nuclei at low energy.  In the high (low) energy range the nucleon (nuclear) degrees of freedom are involved.

The neutrino low energy range is particularly attractive for timely applications, such as the precise calibration of neutrino detectors for the future observation of neutrinos from a core-collapse supernova explosion, or of the diffuse supernova neutrino background.  While electron anti-neutrinos can be detected through proton scattering in water Cherenkov and scintillator detectors,  neutrino-nucleus scattering is necessary for the identification of the electron neutrinos.
Neutrino detectors running or under study exploit nuclei. For example,
a lead-based supernova observatory -- the HALO project -- is now planned at SNOLAB.
It has been shown e.g. in~\cite{Engel:2002hg} that the measurement of neutrino-scattering on lead, in coincidence with (1 or 2) neutrons, has an interesting sensitivity upon the third neutrino mixing angle. 
Large-size detectors, such as  GLACIER or MEMPHYS (UNO, Hyper-K) and LENA, are currently under study~\cite{Autiero:2007zj}. In these observatories the $\nu_e$ of the diffuse supernova neutrino background can be measured through scattering on argon~\cite{Cocco:2004ac}, on oxygen and on carbon~\cite{Volpe:2007qx}. Measurements of neutrino-nucleus cross sections are not only necessary to determine precisely detectors responses but for several other applications.
In~\cite{Haxton:1987bf} an original strategy for the detection of relic galactic supernova neutrinos is proposed, based on the geochemical measurement of $^{97}$Tc in $^{98}$Mo ore.
As pointed out in~\cite{Lazauskas:2009yh} a precise
measurements of the $\nu$-$^{98}$Mo and $\nu$-$^{97}$Tc cross sections is necessary to extract unambiguously the supernova contribution. Neutrino-nucleus cross sections on stable and radioactive nuclei are one of the observables needed to understand stellar nucleosynthesis and, in particular, the r-process  (see e.g.~\cite{Meyer:1998sn,Surman:1998eg}). Unravelling its site is one of the major open questions in nuclear astrophysics, one of the candidates being core-collapse supernova explosions (see e.g.~\cite{Balantekin:2003ip}).

A precise knowledge of the nuclear response to neutrinos is also crucial in order to calibrate some phenomenological ingredients (e.g. knowledge of the forbidden multipoles and their possible quenching) of the microscopic approaches currently used to study neutrinoless double-beta decay~\cite{Ejiri:2003ap,Volpe:2005iy}. 
The observation of neutrinoless double-beta decay would represent a major discovery, from which key information on the CP Majorana phases, on the electron neutrino effective mass and mass hierarchy can be extracted \cite{Elliott:2004hr,Bilenky:2002aw}.
However this requires
reducing the current discrepancies among the half-life predictions.
In this respect a step forward can be made through neutrino-nucleus measurements, combined with other measurements on the candidate emitters (beta-decay, muon capture, charge-exchange reactions and the two-neutrino double beta decay) \cite{Zuber:2005fu}.

The theoretical description of neutrino-nucleus cross sections in the low energy range benefits from a variety of sophisticated models, including Effective Field Theories~\cite{Kubodera:1993rk,Kubodera:2009au}, the Continuum Random-Phase-Approximation (CRPA)~\cite{Kolbe:2003ys,Jachowicz:2002rr}, the Quasi-particle RPA (QRPA)~\cite{Volpe:2000zn} and projected QRPA~\cite{Samana:2008pt}, relativistic RPA~\cite{Paar:2007fi}, the Shell Model (SM)~\cite{Volpe:2000zn,Hayes:1999ew} and the Shell Model in the complex energy plane~\cite{Civitarese:2007ht}. Regardless the high degree of sophistication achieved, important discrepancies remain between the predicted neutrino-nucleus cross sections. For example, in ~\cite{Samana:2008pt} it is shown that the calculations of the energy dependent cross section on $^{56}$Fe differ as much as by factor 3-4 at a given energy, whereas
 the convoluted neutrino cross sections turn to be in agreement with the experimental data\footnote{Note that the experimental uncertainties are in this case at the level of 50\%.~\cite{Zeitnitz:1998qg}}~\cite{Zeitnitz:1998qg}. In ~\cite{Paar:2007fi} the comparison of the flux-averaged neutrino-lead cross sections, associated to supernova neutrinos, show very similar discrepancies.  
 
The realization of precise measurements of the neutrino scattering cross sections for an ensemble of nuclei therefore should help to pin down differences among the models, enabling accurate and reliable description of
 the isospin and spin-isospin nuclear response. This is indeed a mandatory  step for many innovative and timely applications.

\begin{table}[tbp]
\begin{center}
\begin{tabular}{|c|c|c|c|c|c|}
\hline
\multicolumn{1}{|c|}{Facility} &
\multicolumn{1}{|c|}{Power} &
\multicolumn{1}{|c|}{Proton energy} &
\multicolumn{1}{|c|}{Time structure} &
\multicolumn{1}{|c|}{Repetition rate} \\ \hline
LANSCE (USA) & 56 kW & 0.8 GeV & Continuous & N/A \\
ISIS (UK)  & 160 kW & 0.8 GeV & 200 ns & 50 Hz \\
SNS (USA) & $>$ 1 MW & 1 GeV & 380 ns & 60 Hz \\
JSNS (Japan)  & 1 MW & 3 GeV & 1 $\mu$s & 25 Hz \\
SPL (CERN) & 4 MW & 3.5 GeV & 0.76 ms & 50 Hz \\
ESS (Sweden) & 5 MW & 1.3 GeV & 2 ms (1.4 $\mu$s) & 17 Hz (50 Hz)
\\ \hline
\end{tabular}%
\end{center}
\par
\vskip 0.5cm
\caption{Comparison of characteristics of the past, present, and future Spallation Source Facilities in different regions of the world\label{t:facilities}.}
\end{table}

Coherent neutrino-nucleus scattering is another important phenomenon that has never been measured.
Such an experiment is useful as a test for the Standard Model but also as a technological
advancement for supernova neutrino detectors and dark matter searches.  The cross section in the 50 MeV energy range can be as large as 10$^{-39}$cm$^{2}$. The CLEAR experiment is now planned at SNS to measure coherent neutrino-nucleus scattering~\cite{Vergados:2009ei,Scholberg:2009ha}. Furthermore, neutrino capture on radioactive nuclei would open a completely new window on the Universe since this process could be used to observe the cosmological neutrino background. This exciting possibility has been pointed out in~\cite{Cocco:2007za} and further investigated in~\cite{Lazauskas:2007da}.
Note that, since this background has a temperature of 1.96 K, the impinging neutrino energy is so low, that one can calibrate the corresponding cross sections using the inverse process (the beta-decay of the nuclei of interest).

Detailed investigations of the physics scope of spallation source facilities as far as neutrino physics, neutrino astrophysics, tests of the Standard Model are concerned, can be found in the literature (see e.g.~\cite{Avignone:2003ep}). Ref.~\cite{sns} is a proposal made for the SNS facility presenting also an in-depth study of the possible backgrounds and of the detectors' design. In this paper, motivated by the recently approved European Spallation Source facility, we present new predictions for neutrino-nucleus, neutrino-proton and electron scattering rates. We emphasize general aspects, relevant for the cross section measurements and for low energy tests of the Standard Model, in particular a Weinberg angle measurement and Lepton-Flavour Violation (LFV) searches.

\section{Results on expected rates}
We assume here a neutrino flux at the source of 10$^{15}~\nu_e$/s and a fully efficient 1 ton cubic detector.
At this early stage, we present only very general estimates for the expected
neutrino event rates, neglecting statistical and systematic errors coming from possible backgrounds. Therefore,  
our numbers of events can be easily scaled, for any neutrino production rate and experiment running time.

\subsubsection*{Nuclear excitations:}
\begin{table}[tbp]
\begin{center}
\begin{tabular}{|l|ccc|cc|}\hline
& 10 & 20 & 50 & $\rho $ (g/cm$^{3})$ &$<\sigma> _{DAR}$ \\ \hline\hline
$^{12}$C (in C$_{16}$H$_{18}$) & 1470 & 384 & 63 & 0.992 & $\approx$0.14\cite{Zeitnitz:1998qg,bib:nuC12} \\
$^{16}$O (in water) & 998 & 261 & 43 & 1. & 0.131~\cite{Lazauskas:2007bs} \\
$^{40}$Ar & 8860 & 2310 & 380 & 1.43 & 2.56~\cite{Kolbe:2003ys} \\
$^{56}$Fe  & 9100 & 2330 & 377 & 7.87 & 3.53~\cite{Lazauskas:2007bs} \\
$^{100}$Mo  & 17300 & 4420 & 716 & 10.28 & 11.95~\cite{Lazauskas:2007bs} \\ \hline
$^{208}$Pb  & 34500 & 8820 & 1430 & 11.34 & 49.6~\cite{Lazauskas:2007bs} \\
$^{208}$Pb + 1n & 1630 & 4180 & 677 &  & 23.5~\cite{Engel:2002hg} \\
$^{208}$Pb + 2n & 9420 & 1140 & 390 &  & 13.5~\cite{Engel:2002hg} \\ \hline\hline
\end{tabular}

\end{center}
\caption{Results on the number of events at a neutrino experiment based at a spallation source facility.
The events are calculated assuming 10$^{15}~\nu_e$/s, in a year ($3~10^{7}$~s),
with a fully efficient 1 ton cubic detector. The columns correspond to the considered targets
(first column), the rates at different distances d (meters) from the source, the material density (fourth column) and the flux-averaged cross sections in units $10^{-40}$~cm$^2$ (last column).  }
\label{t:rates1}
\end{table}

\begin{figure}[t]
\begin{center}
   \includegraphics[scale=0.7,angle=0]{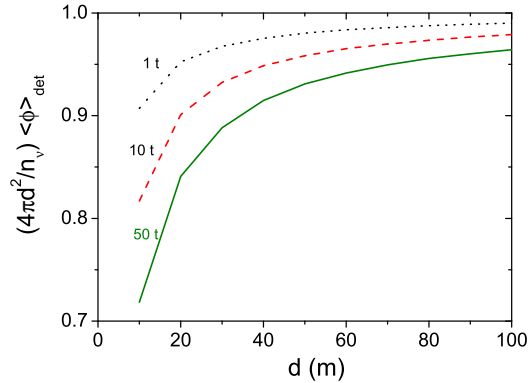}
\end{center}
   \caption{Effect of the finite size of the detector :  Ratio of the flux integrated over the detector volume, over the flux at distance d, as a function of the distance from the spallation source. }
   \label{fig:w}
\end{figure}

Neutrino-nucleus charged-current event rates
are given in Table \ref{t:rates1}, assuming different target nuclei. The detector
location is varied, its front being set
at 10, 20 and 50 meters  from the source. It is obvious that the detector should be placed as close
as possible to the source in order to increase its response,
the neutrino flux decreasing as the inverse distance squared.
At such short distances the detector finite size also counts, for the number of events.
Figure~\ref{fig:w} shows this effect as a function of the distance from the source.

We present rates for nuclei that can be found in detectors based on various technologies, namely carbon (detectors made of liquid scintillators, e.g. $C_{16}H_{18}$), oxygen (for water Cherenkov), argon, iron, molybdenum and lead.
The corresponding neutrino-nucleus cross sections, averaged over the decay-at-rest (DAR) muon flux (Michel spectrum, Figure \ref{fig:flux}), are chosen as follows. We take the experimental flux-averaged cross sections
for $^{12}$C \cite{Zeitnitz:1998qg,bib:nuC12}. Our
QRPA calculations with the Skyrme interaction are used for $^{16}$O, $^{56}$Fe, $^{100}$Mo and  $^{208}$Pb from Ref.~\cite{Lazauskas:2007bs}. The $^{208}$Pb cross sections with neutron(s) emission are from Ref.~\cite{Lazauskas:2007bs}. CRPA calculations
of the $^{40}$Ar cross sections are taken from ~\cite{Kolbe:2003ys}.
The electron anti-neutrino-proton cross section has been calculated,
up to all orders, using formulae derived by  Ya. I. Azimov and V. M. Shekhter~\cite{bib:azim}.

\begin{table}[tbp]
\begin{center}
\begin{tabular}{|c|cccc|cccc|}
\hline
\multicolumn{1}{|c|}{$~^{A}$X~} &
\multicolumn{1}{c}{$~0^{+}~$} & \multicolumn{1}{c}{$~1^{+}~$} &
\multicolumn{1}{c}{$~2^{+}~$} & \multicolumn{1}{c|}{$~3^{+}~$} &
\multicolumn{1}{c}{$~0^{-}~$} & \multicolumn{1}{c}{$~1^{-}~$} &
\multicolumn{1}{c}{$~2^{-}~$} & \multicolumn{1}{c|}{$~3^{-}~$} \\ \hline
$^{16}$O & 0.23 & 1.38 & 0.70 & 0.74 & 0.79 & 42.3 & 53.7 & 0.1 \\
$^{56}$Fe & 13.0 & 66.7 & 1.09 & 1.12 & 0.24 & 9.45 & 8.27 & 0.06 \\
$^{100}$Mo &  16.7 & 53.3 & 2.12 & 2.08 & 0.30 & 12.2 & 12.6 & 0.29 \\
$^{208}$Pb & 12.9 & 43.3 & 4.95 & 4.76 & .317 & 14.1 & 17.1 & 1.14 \\ \hline

\end{tabular}%
\end{center}
\par
\vskip 0.5cm
\caption{Fraction (in $\%$) of the flux-averaged cross section associated to
states of a given multipolarity, with respect to the total flux-averaged
cross section, i.e. $\langle \protect\sigma \rangle_{J^{\protect\pi%
}}/\langle \protect\sigma \rangle_{tot}$. Results are given for
all positive and negative states, having total angular momentum $J$
between 0 and 5. The first column gives the considered nucleus~\cite{Lazauskas:2007bs}.}\label{t:multi}.
\end{table}

\begin{figure}[t]
\vspace{.6cm}
\begin{center}
\centerline{\includegraphics[scale=0.5,angle=0]{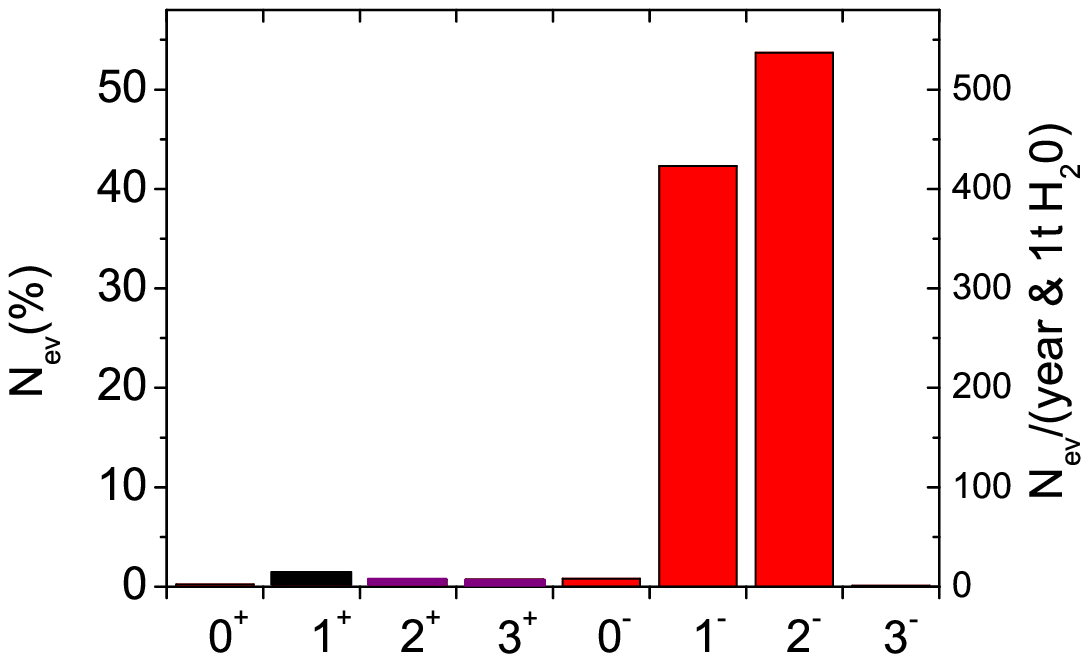}\hspace{.2cm}
\includegraphics[scale=0.5,angle=0]{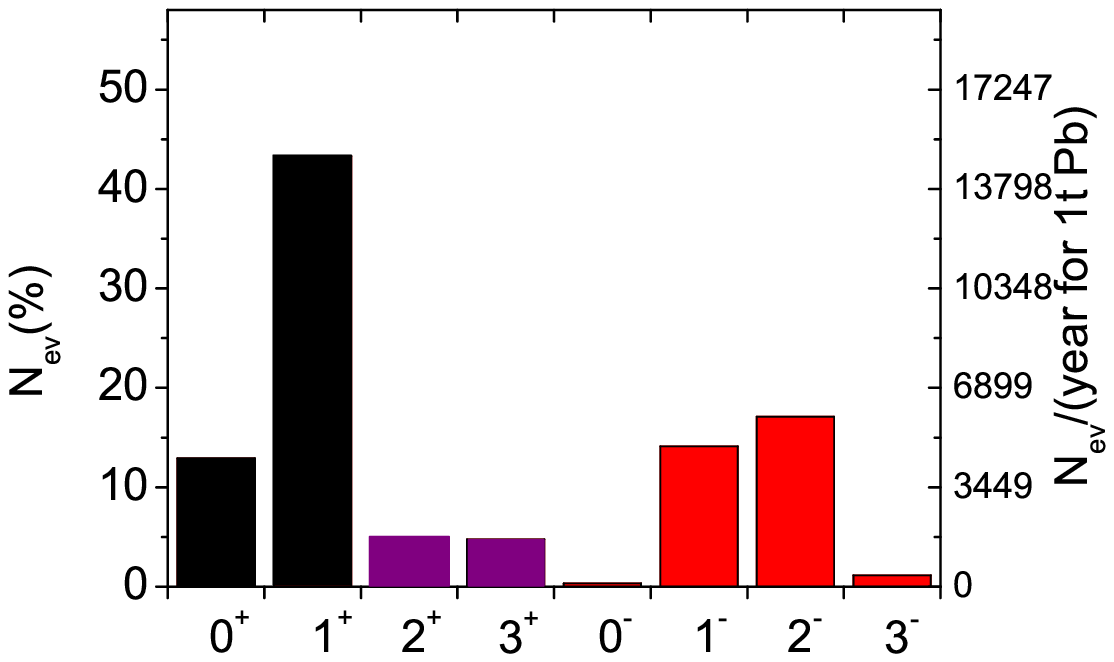}}
\end{center}
\caption{Contributions of the positive and negative parity states, in percentage, to the number of events for oxygen (left) and lead (right figure) nuclei, as examples.}
\label{fig:multi}
\end{figure}

As can be seen from Table \ref{t:rates1} convolved DAR cross sections as well as the event rates are lowest for
carbon and oxygen
nuclei. This is mainly due to the high reaction threshold but, for oxygen, also due to the closed shell structure.
The most interesting event rates are obviously from
detectors based on heavy nuclei.

\begin{table}[tbp]
\begin{center}
\begin{tabular}{|l|ccc|cc|}\hline
& 10 & 20 & 50 & $\rho $ (g/cm$^{3})$ & $<\sigma >_{DAR}$ \\ \hline\hline
$\nu_e $-e  & 230 (223) & 60 (58) & 10 & 1.0 (0.992) & 2.99$\cdot 10^{-3}$ \\
$\nu_{\mu} $-e & 35 (34) & 9.15 (9.0) & 1.52 (1.5) & 1.0 (0.992) & 4.58$\cdot 10^{-4}$ \\
$\overline{\nu}_{\mu} $-e  & 38 (37) & 9.9 (9.6) & 1.6 (1.5) & 1.0 (0.992) & 4.93$\cdot 10^{-4}$
\\\hline
\end{tabular}
\end{center}
\caption{Same as Table \ref{t:rates1} (anti)neutrino scattering on electrons in water. The values given in parenthesis correspond to the results for electrons in C$_{16}$H$_{18}$ as another example.}
\label{t:rates2}
\end{table}

In order to understand the nuclear structure response, involved in the DAR flux-averaged
cross sections, it is necessary to identify the different multipole contributions.
Note that a precise knowledge of the nuclear transition matrix elements of states of high multipolarity is crucial
to better constrain neutrino-less double beta-decay predictions \cite{Volpe:2005iy}.
Note that part of the forbidden contributions are usually located at higher energies and can be best studied when one can vary the
average energy of the impinging neutrinos.
In this respect low energy beta-beams constitute a powerful tool since the average neutrino energy can be increased by boosting the ions at higher energies.
On the other hand DAR neutrino fluxes from Spallation Sources have
much higher intensities than the ones that will be achieved by low energy beta-beams.
In \cite{McLaughlin:2004va} a comparative analysis of the multipole contributions to the neutrino-lead cross section at spallation sources and low energy beta-beams is made, showing that measurements at these facilities are complementary. Ref.\cite{Serreau:2004kx} has first investigated neutrino-nucleus cross section measurements at a low energy beta-beam facility while the multipole decomposition of the corresponding flux-averaged cross sections are given in \cite{Lazauskas:2007bs}.
In Figure~\ref{fig:multi} and Table~\ref{t:multi} we
show the decomposition of the total cross section
over the states of different multipolarity, that contribute to the cross sections.
For a closed shell nucleus as oxygen, the
forbidden $1^{-},~2^{-}$ excitations dominate the cross section.
The case of oxygen is exceptional: the allowed (Gamow-Teller and Fermi) transitions
accounts  for less than $2\%$ of the total cross section. In most cases allowed excitations are
the dominant contribution; while the forbidden one can be significant. For example it is at the level
of $30\%$ for $^{100}$Mo and $^{208}$Pb.

\subsubsection*{A Weinberg angle measurement:}
Neutrino detectors can also be used to study neutrino-electron scattering.
The expected number for $\nu_{e,\mu}$-e  and $\overline{\nu}_{\mu}$-e scattering
are given in Table~\ref{t:rates2}. Although the $\nu$-e cross section is much
weaker compared to those on nuclei, such events can be separated in the detector by
considering forward scattering.
If the systematic errors are kept sufficiently low, one can use this measurement to extract
non-standard contributions to the weak interaction (e.g. FCNC effects)
and perform a precise measurement of the Weinberg angle at low momentum transfer.
The current best value for $\nu_e-e$ scattering is $\sigma = 10.1 \pm 1.1 (stat)  \pm 1.0 (sys) E_{\nu_e} \times 10^{-45} cm^{2}$ \cite{Auerbach:2001wg}. Note that a possible measurement with low energy beta-beams can determine the Weinberg angle with a precision of 10 $\%$ if the systematic errors are kept below $10\%$ \cite{Balantekin:2005md}.

\begin{figure}[t]
\vspace{.6cm}
\begin{center}
\includegraphics[scale=1.,angle=0]{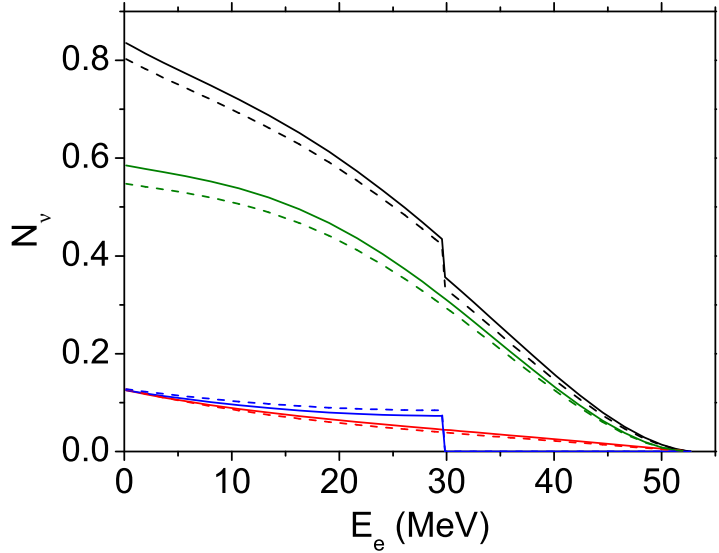}
\end{center}
\caption{Energy distributions of the $\nu$-e events, in relative units, for a Weinbrg angle $sin^2\theta_W=0.21$ (dashed),
$0.23$ (full lines). The upper (two) curves represent the total number of electron events,
the next lines correspond to: the $\nu_e$-e events, followed by the $\nu_{\mu}$-e and the $\overline{\nu}_{\mu}$-e ones.}
\label{fig:nu_e}
\end{figure}

The time-structure of the beam pulse can be crucial for the accurate determination
of the  Weinberg angle at spallation sources. Indeed, positive pion ($\pi^+$) decay produce neutrinos of $\overline{\nu}_{\mu}$,
$\nu_e$ and  $\nu_{\mu}$ flavors, whose scattering on electrons have a very different reaction cross section
dependence on the Weinberg angle, as can be seen in Figure~\ref{fig:nu_e}. The $\nu_e$-e and  $\overline{\nu}_{\mu}$-e cross sections increase for an increasing value of  the Weinberg angle, while $\nu_{\mu}$-e
cross section decreases.

Let us define the quantity :
\begin{eqnarray}
S=|\frac{dN(sin^2\theta_W)}{dsin^2\theta_W}| \frac{sin^2\theta_W}{N(sin^2\theta_W)},
\end{eqnarray}
that we will call the sensitivity, where $N$ stands for the experimental observable to be used in determining the Weinberg angle $sin^2\theta_W$.
It turns out that the measurement of the total number of neutrino-electron events
has a sensitivity of $S=0.5$ only.

If the pulse length is made much shorter than the muon life-time, the $\nu_{\mu}$-e events can be separated
from $\overline{\nu}_{\mu}$-e and  $\nu_{e}$-e  ones. This enables one to study the quantity -- as proposed for the SNS facility
in \cite{Avignone:2003ep,sns} --
\begin{eqnarray}
O=\frac{N({\nu}_{\mu}-e)}{N(\overline{\nu}_{\mu}-e)+N(\nu_{e}-e)},
\end{eqnarray}%
whose sensitivity to the Weinberg angle is as much as $S=1.68$.

If however one dispose of detector with good energy resolution the determination of the Weinberg angle can be
improved even in the case of long pulses. To this aim one has to determine the size of the jump in the electron energy
distribution at $E_e=29.7$ MeV, the end point of the  ${\nu}_{\mu}$-e events. This can be done by
fitting separately the electron energy distributions above and below $29.7$ MeV. The jump height
has a sensitivity S to the Weinberg angle as large as 2.
The measurement made by the LSND collaboration, after angular cuts, has left about 230 events
identified as $\nu$-e scattering. With a only 5 ton detector at ESS e.g. one could expect in a running time of 1 year
almost 5 times more events.

\begin{table}[tbp]
\begin{center}
\begin{tabular}{|l|ccc|cc|}\hline
& 10 & 20 & 50 & $\rho $ (g/cm$^{3})$ & $<\sigma> _{DAR}$ ($10^{40}$~cm$^2$) \\ \hline\hline
$\overline{\nu }$-H (water) & 11300 & 2950 & 490 & 1.0 & 0.738~\cite{bib:azim} \\
$\overline{\nu }$-H (C$_{16}$H$_{18}$) & 8720 & 2280 & 374 & 0.992 &
0.738~\cite{bib:azim}\\\hline
\end{tabular}
\end{center}
\caption{Same as Table \ref{t:rates1} but for electron anti-neutrino scattering on protons. }
\label{t:rates3}
\end{table}

\subsubsection*{Lepton-Flavour-Violation:} Finally let us discuss the case of $\overline{\nu}_e$ scattering. If the amount of anti-neutrinos
is as low as expected at the SNS facility, namely $10^{-5}$, one could use this very pure source of neutrinos to search for rare processes
such as the Lepton-Flavour-Violating decay $\mu^+\rightarrow\e^+\overline{\nu}_e\nu_\mu$.
This search has been performed, e.g. by the KARMEN Collaboration, giving the current best limit for the branching ratio of $0.9 \times 10^{-3}$ \cite{Armbruster:2003pq}. Table \ref{t:rates3} gives the expected $\overline{\nu}_e$-p scattering rates assuming that as many $\overline{\nu}_e$ as
$10^{15}\overline{\nu}_e$/s are produced, these numbers must be scaled according to the suppression factor, that depends on the specific spallation source considered.
This experiment requires a precise identification of the associated events, measuring the neutron through
neutron capture on Gadolinium or on protons.

\section{Conclusions}
New Spallation Source facilities  are under construction, planned or under study. They will produce an intense flux of neutrinos.
If these facilities are designed to accommodate a neutrino detector(s) close to the source, they will offer an unique
 opportunity to perform low energy neutrino scattering experiments.
 Neutrino detectors with different active material can be used to perform neutrino nucleus cross section measurements that have a variety of innovative and timely applications going from nuclear physics to neutrino astrophysics.
Neutrino-nucleus scattering data would create conditions for the breakthrough of the theoretical models, permitting to pin  down some of their phenomenological ingredients. Besides,
physics beyond the standard model can be tested through the searches for non-standard contributions to neutrino electron scattering; while
the identification of electron anti-neutrino scattering on protons can be used to improve established limits on rare lepton
 flavor violating decays, in particular
$\mu^+ \rightarrow \e^+ + \bar{\nu}_e + {\nu}_{\mu}$. We have presented new predictions for these processes and emphasized some general aspects, hoping one day 
such measurements to be realized.

\vspace{.3cm}
\noindent
{\bf Acknowledgements}
The authors thank G. Drexlin and M. Mezzetto for useful discussions.

\end{document}